# Control of rectifying and resistive switching behavior in BiFeO$_3$ thin films

Yao Shuai,[1,2] Shengqiang Zhou,[1] Chuangui Wu,[2] Wanli Zhang,[2] Danilo Bürger,[1] Stefan Slesazeck,[3] Thomas Mikolajick,[3] Manfred Helm,[1] and Heidemarie Schmidt[1]

[1]*Institute of Ion Beam Physics and Materials Research,*

*Helmholtz-Zentrum Dresden-Rossendorf, P. O. Box 510119, Dresden 01314, Germany*

[2]*State Key Laboratory of Electronic Thin Films and Integrated Devices,*

*University of Electronic Science and Technology of China, Chengdu 610054, China*

[3]*Namlab gGmbH, Nöthnitzer Strasse 64, 01187 Dresden, Germany*

**Abstract:** BiFeO$_3$ thin films have been grown on Pt/Ti/SiO$_2$/Si substrates with pulsed laser deposition using Au as the top electrode. The resistive switching property of the Au/BiFeO$_3$/Pt stack has been significantly improved by carefully tuning the oxygen pressure during the growth, and a large switching ratio of ~4500 has been achieved. The deposition pressure modifies the concentration of oxygen vacancies and the rectifying behavior of the Au/BiFeO$_3$ junction, and consequently influences the resistive switching behavior of the whole stack. The switching takes place homogeneously over the entire electrode, and shows a long-term retention.

Since semiconductor nonvolatile memories, for example, flash memories, are reaching the downscaling limit, it is becoming urgent to search for new materials, which can replace memories based on the conventional silicon CMOS technology. So far resistive random access memory (RRAM) materials are the most promising candidates.[1-6] Resistive switching has been reported in a large number of compounds, particularly in oxides. However, the observed switching behavior varies significantly even in the same type of material.[1-7]

Resistive switching has been reported recently in BiFeO3 thin films,[3-8] which have been intensively investigated concerning their multiferroic properties. It is therefore interesting to study the resistive switching behavior in BiFeO3 thin films, which can help to guide the future design of multifunctional devices combining ferroelectricity, ferromagnetism, and charge conduction. The previously reported resistive switching in BFO thin films mostly shows symmetric current-voltage (I-V) characteristics[3,4] and suffers possible negative effects from unchecked sneak currents.[9,10] Switchable diode effects combined with resistive switching have been reported for BFO single crystal[8] or BFO thin films epitaxially grown on single crystal substrates,[6,7] however, the ON/OFF ratio is only ~100, moreover, the complexity of the fabrication process of single crystal BFO or the high cost of single crystal substrates hamper their practical industrial applications.

We previously observed the interface-related resistive switching in polycrystalline $BiFeO_3$ (BFO) thin films deposited on low cost Pt/Ti/SiO2/Si substrates,[11] which also show a rectifying behavior. However, the precise control of the resistive switching property together with the rectifying behavior during the thin film fabrication is still lacking and needs to be further studied. In the present work, the influence of the deposition pressure on the switching behavior of BFO thin films has been studied, the resistive switching and the rectifying behavior in BFO thin films were found to be strongly dependent on the oxygen vacancy concentration, which is very sensitive to the deposition pressure of the BFO thin films. Within a narrow growth window regarding the oxygen pressure, a significant enhancement of resistive switching and rectifying behavior have been achieved.

The BFO thin films were deposited onto Pt/Ti/SiO$_2$/Si substrates using pulsed laser deposition (PLD). The substrate temperature was kept at 700 °C, while the oxygen partial pressure was varied from 6 to 40 mTorr, and the BFO thin films are, respectively, labeled as BFO-6, BFO-8, BFO-10, BFO-13 and BFO-40, where the numbers represent the oxygen pressure in units of mTorr. For electric measurements, Au top electrodes with a size ranging between 0.03 and 1 mm$^2$ were deposited by RF sputtering using a metal shadow mask. I-V curves and pulsed voltage measurements were done with a Keithley 2400 source meter.

Fig. 1(a) shows the I-V curves of those BFO thin films deposited under different oxygen partial pressures. The positive bias is applied to the top Au electrode, while the bottom Pt electrode is grounded. Before the I-V measurement, all the samples are poled by a -8 V voltage pulse. The sequence of the voltage for the I-V measurement is as follows: 0V → +10V → 0V → -10V → 0V. As illustrated in Fig. 1(a), sample BFO-10 exhibits the most obvious hysteretic I-V curve (black solid circle). It is at the high resistance state (HRS, range (1)) after poling. After the voltage reaches the maximum positive value (+10V), it turns back to sweep downward, and a hysteretic I-V curve starts to appear, which indicates that the thin film has been set to the low resistance state (LRS, range(2)). The thin film can be reset to HRS by applying a negative bias, as shown by range (3) and (4) in Fig. 1(a). The resistance ratio between HRS and LRS ($R_{HRS}/R_{LRS}$) amounts to ~4500 at +2V and is large enough for practical applications. The switching behavior is significantly influenced by the deposition pressure [Fig. 1(a)]. For example, the $R_{HRS}/R_{LRS}$ ratio decreases dramatically to ~110 (104) in sample BFO-13 (BFO-8) by slightly increasing (decreasing) the deposition pressure. By further increasing the deposition pressure, the resistive switching nearly disappeared in sample BFO-40. On the other hand, the resistive switching was also suppressed considerably by decreasing the deposition pressure further, i.e. $R_{HRS}/R_{LRS}$ is only ~12 in sample BFO-6. Besides, the rectification factor (RF) is strongly dependent on the deposition pressure as well, the largest RF is also obtained in BFO-10, which shows the largest $R_{HRS}/R_{LRS}$. The $R_{HRS}/R_{LRS}$ and RF are plotted as a function of the deposition pressure as shown in Fig. 1(b). Large $R_{HRS}/R_{LRS}$ and RF can only be achieved in a very narrow window, indicating that the deposition pressure is a critical factor for obtaining the large resistive switching ratio and rectification factor in BFO thin films.

Now let us turn to understand the large difference concerning the switching behavior of the BFO films grown at different oxygen pressures. It can be seen that the BFO-8, BFO-10, and BFO-13 samples with a relatively large $R_{HRS}/R_{LRS}$ ratio reveal rectifying I-V characteristics [Fig.1(a)], while the BFO-6 and BFO-40 samples with a much lower $R_{HRS}/R_{LRS}$ ratio show relatively symmetric I-V characteristics. Therefore, it is obvious that the rectifying top contacts in BFO-8, BFO-10 and BFO-13 play an important role in the resistive switching behavior. Because oxygen vacancies are always present in perovskite oxides, it is reasonable to regard the BFO thin films as n-type semiconductors.[7] When Au or Pt is in contact to the n-type BFO, a Schottky contact should be formed, because both of them have a high work function. However, the BFO-10 (6, 8 or 13) /Pt interface is an Ohmic-like contact which can be deduced from the large current in the positive bias range [Fig. 1(a)]. This is likely due to the interdiffusion at the Pt bottom contact caused by the high growth temperature, which prevents a Schottky contact formation at the BFO/Pt interface. On the other hand, the rectifying current in the negative bias range [Fig. 1(a)] reveals a Schottky contact formation at the Au/BFO-10 (8, 13 or 40) interface. Detailed discussion about the transport characteristic in this structure can be found in our previous report.[11] The result in Fig. 1(a) confirms that the resistive switching in BFO-10 (8 or 13) depends on this asymmetric contact type between the electrodes and the thin film, which can be controlled by carefully tuning the oxygen partial pressure during the thin film growth.

Note that before the cycling of the voltage, a poling process was applied using a -8 V pulse. We have previously given a possible switching mechanism in this asymmetric structure, which is due to an electron trapping and detrapping effect at those electron trapping sites inside the BFO.[11] The poling with a negative bias can remove the trapped electrons and extend the depletion region in the pristine thin film, significantly increasing the resistance of the structure. Therefore, a large switching ratio can be observed in the following I-V measurement. Without poling and I-V hysteresis history, the virgin BFO-10 shows a minor I-V hysteresis at the beginning as illustrated by the black dashed line in Fig. 1(a). The observed poling process is different from the frequently reported forming process, because there was no conductive filament formed and the resistance of the pristine sample is increased instead of being decreased.[12,13]

The resistive switching with a large $R_{HRS}/R_{LRS}$ ratio is absent in samples BFO-6 and BFO-40. The Ohmic-like contacts at both interfaces of sample BFO-6 allow the electron injection at both voltage polarities, therefore, BFO-6 keeps at LRS due to the absence of depletion region. On the other hand, sample BFO-40 has Schottky contacts at both interfaces. Because one of the two Schottky contacts is always reversely biased, BFO-40 is always at HRS and not strongly resistively switchable.

In addition to the contact rectification also the concentration of oxygen vacancies influences the resistive switching behavior of BFO thin films. The slight increase (decrease) of growth pressure from 10 to 13 (8) mTorr considerably decreased the $R_{HRS}/R_{LRS}$ (Fig. 1). It indicates that even if the structure has an asymmetric barrier geometry, the density of trapping centers inside the BFO thin film is still a critical factor for the switching. To obtain a large switching ratio in the BFO thin films, the growth pressure must be carefully controlled not only to form just one rectifying contact, but also to get a suitable concentration of oxygen vacancies.

Note that the resistive switching of sample BFO-10 takes place homogeneously over the entire electrode. We measured the I-V curves of sample BFO-10 using different electrode sizes (Fig. 2). The current shows a linear dependence on the electrode area (inset of Fig. 2). That is similar with the observation for a Nb:SrTiO$_3$ memory cell, which also shows a linear relationship between the electrode size and current.[14]

To confirm the applicability, we examined the retention of the resistive switching of sample BFO-10 by pulsed voltage measurements. As shown in Fig. 3, HRS and LRS were realized by writing pulses with amplitude of -9 V and +9 V, respectively. This writing process is followed by a reading pulse of +2 V, which is repeated for several thousands of times for resistance detection. A large $R_{HRS}/R_{LRS}$ above $10^3$ can be observed in Fig. 3, which is in good agreement with the hysteretic I-V curve in Fig. 1. Note that the smaller reading pulse of +2 V does not change the $R_{HRS}/R_{LRS}$ obviously even it is repeated continuously, which hints towards non-destructive read-out characteristics. The HRS is stable, while the LRS shows a fast decay at the beginning and becomes stable after 3000 seconds. However, neither of the HRS and LRS shows an obvious change after three months (inset of Fig. 3), revealing a long term retention.

It has to be pointed out that, a pure electronic effect involving electron trapping and detrapping may not fully explain the long term retention characteristic observed in our Au/BFO/Pt structure. Other groups have understood the resistive switching in BFO thin films by means of the polarization switching.[4,5] However, the P-E loops of our BFO thin films are unsaturated and the coercive voltage is much larger, therefore, the voltage for the I-V measurement is not able to sufficiently switch the polarization of the BFO thin films. On the other hand, Muenstermann *et al.*[1] suggested that the switching takes place near the bottom interface due to the migration of oxygen vacancies. However, that is not applicable in our case either. Because of the asymmetric contact barrier, the external electric field mainly drops across the depletion region at the top Au/BFO interface, while the electric field strength near the bottom BFO/Pt interface is small, thus the migration of oxygen vacancy near the bottom interface can nearly be neglected. Therefore, the nonvolatile resistive switching with long term retention and the underlying switching mechanism in our Au/BFO/Pt structure needs to be further investigated.

In summary, $BiFeO_3$ thin films have been grown onto $Pt/Ti/SiO_2/Si$ substrates under different oxygen partial pressures. The resistive switching property and rectifying behavior can be controlled by tuning the growth pressure, which determines the concentration of oxygen vacancies in the $BiFeO_3$ thin films and the contact type at the interface. The asymmetric contact geometry has been demonstrated to be crucial for the resistive switching, and the concentration of oxygen vacancies has also to be carefully controlled to achieve a large switching ratio. The structure shows a long-term retention, which is proved by the pulsed voltage measurement.

Y. S. would like to thank the China Scholarship Council (grant number: 2009607011). S. Z. acknowledges the funding by the Helmholtz-Gemeinschaft Deutscher Forschungszentren (HGF-VH-NG-713). D. B. and H. S. thank the financial support from the Bundesministerium für Bildung und Forschung (BMBF grant number: 13N10144).

**FIGURE CAPTIONS**

Fig. 1. (a) I-V curves of BFO thin films grown at different oxygen partial pressures. The minor IV hysteresis of the virgin BFO-10 at the beginning is also shown (black dashed line). (b) Resistance ratio between HRS and LRS ($R_{HRS}/R_{LRS}$) at +2 V and rectification factor at $\pm 10$ V as a function of deposition pressure. In a small growth window (shaded area) resistance ratio is larger than 100.

Fig. 2. I-V curves of BFO-10 using different electrode sizes. The inset shows a linear relationship between the current and the electrode area.

Fig. 3. Pulsed voltage measurement on BFO-10 (black) after the structure preparation and (red) after three months. The HRS and LRS are induced by +9 V and -9 V pulses, respectively, and the resistance of both states is read out by a +2 V pulse.

Fig. 1(a)

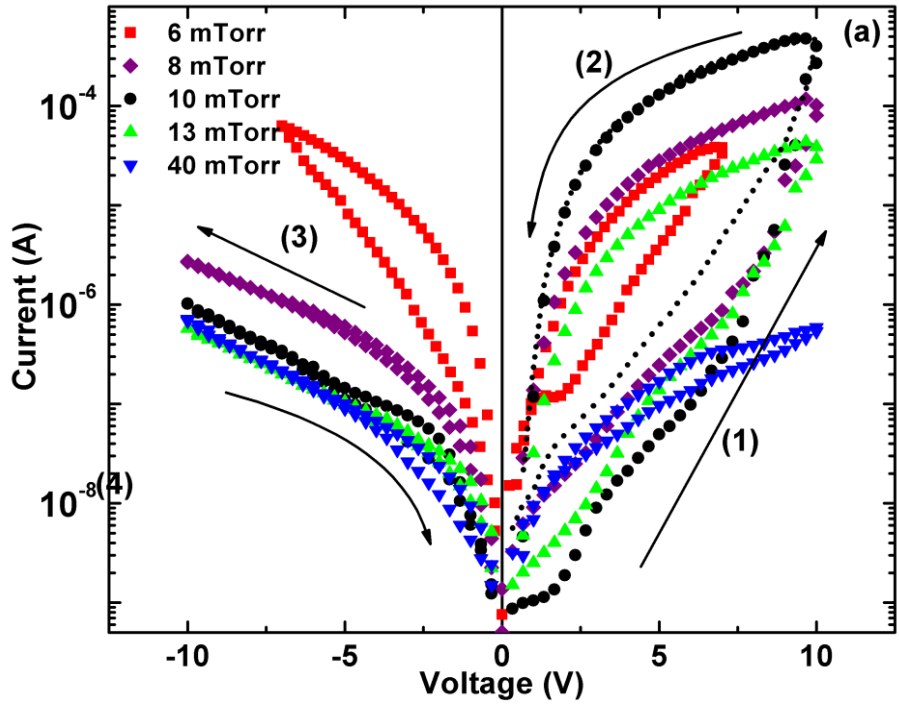

Fig. 1(b)

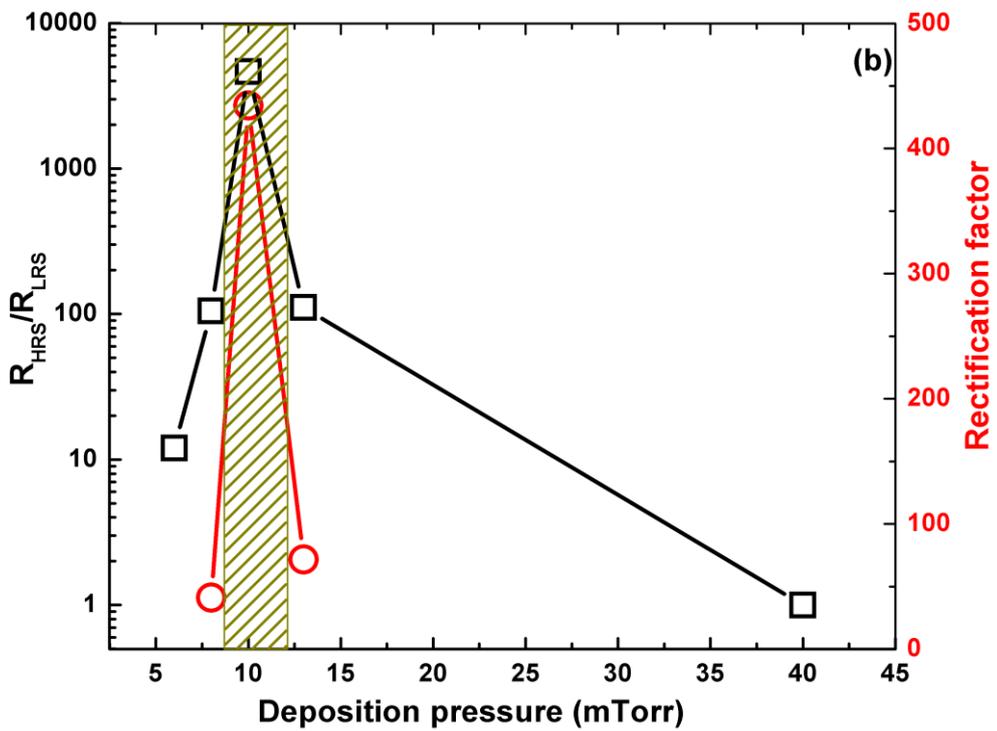

Fig. 2

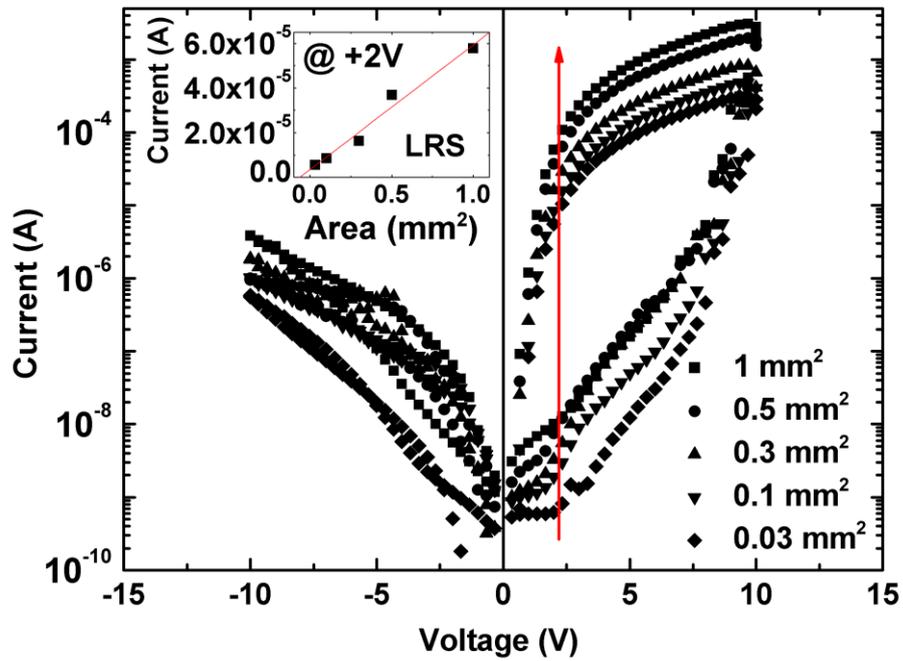

Fig. 3

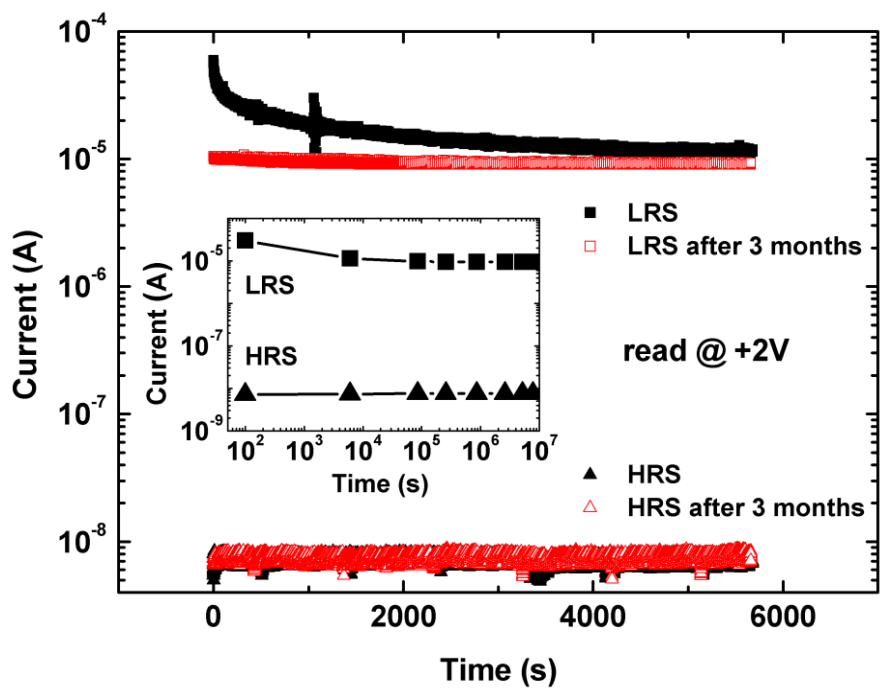